\documentclass[12pt]{article}
\usepackage[T1]{fontenc}
\usepackage{graphicx}

\newcommand{\interlinia}{}

\title{Flexoelectric effect in biaxial nematics}

\author{A. Kapanowski \\
{\em Institute of Physics, Jagiellonian University,}\\
{\em ulica Reymonta 4, 30-059 Cracow, Poland}  }

\begin{document}
\maketitle

\begin{abstract}
\interlinia
The flexoelectric (FE) effect provides a linear coupling between 
electric polarization and orientational deformation in liquid crystals.
It influences many electrooptical phenomena and it is used in some
bistable nematic devices.
A statistical theory of dipole FE polarization in biaxial nematic 
liquid crystals is used to calculate temperature dependence
of order parameters, elastic constants, and FE coefficients.
The splitting of the two Meyer FE coefficients and the appearance
of new FE coefficients is obtained at the uniaxial-biaxial
nematic transition. The ordering of the splited FE coefficients
corresponds to the ordering of the splited elastic constants.
\end{abstract}

\interlinia

\section{Introduction}

The biaxial nematic liquid crystals were first predicted by Freiser
\cite{[1970_Freiser]}, who showed that molecules with shapes
that deviate from cylindrical symmetry could possess a nematic
phase with three distinct optical axes.
A biaxial nematic phase was first observed in a lyotropic
mixture by Yu and Saupe \cite{[1980_Yu_Saupe]}
in 1980 but the existence of a thermotropic biaxial system
was not certain for many years.
Several reports appeared in 2004 on thermotropic liquid crystals
formed by side-chain polymers 
\cite{[2004_Severing_Saalwachter]},
bent-core molecules
\cite{[2004_Madsen],[2004_Acharya]},
and tetrapodes 
\cite{[2004_Merkel]}.
The number of new biaxial systems is constantly growing.

Many theoretical papers 
\cite{[2005_Longa_Grzybowski]}
and computer simulations
\cite{[2008_Berardi]}
show that the molecular shape and
pair interaction biaxiality are important for the biaxial phase
to exist. However, very often real systems favour packing
in the smectic or crystalline biaxial phases.
It is a challenge for the theory to find factors responsible
for absolute stability of the biaxial nematic phase.
It was shown that fluctuations in molecular shape
can influence the biaxial nematic phase stability
\cite{[2007_Longa]}.
The motivation for this search ranges from purely academic interest
to the potential usage of these materials in faster displays,
where in principle the commutation of the secondary director
should give lower response times compared to the conventional
twisted nematic and ferroelectric smectic devices.

A static electric field imposed on a nematic liquid crystal
have many physical effects, but the most important are
two of them. One is connected with the anisotropy of the dielectric
constant. The second effect is the appearance of the spontaneous
polarization in a deformed liquid crystal; this is called
the flexoelectric effect. Conversely, an electric
field may induce distortions in the bulk.
In 1969 Meyer showed that it is a steric effect due to the 
shape asymmetry of polar molecules
\cite{[1969_Meyer]}.
In case of nonpolar molecules the FE effect originates from
a gradient of quadrupole moment density
\cite{[1977_Prost_Marcerou]}.

The two FE coefficients were introduced by Meyer for splay and bend 
distortions of the uniaxial nematic phase
\cite{[1969_Meyer]}.
Recently, a statistical theory for the dipole FE polarization
was derived in the case of the biaxial nematic phase 
composed of the $C_{2v}$ molecules 
\cite{[2008_Kapanowski]}. 
There are six splay-bend deformations of the biaxial nematic phase 
and thus six FE coefficients are defined, 
but only five of them are independent. 
General microscopic expressions for the FE coefficients involve 
the one-particle distribution function and the potential energy 
of two-body short-ranged interactions. 

The FE effect has a large influence on many phenomena in
liquid crystals: electrooptical phenomena and defect formation, for instance. 
It plays a key role in some device applications.
Flexoelectric switching is important in bistable displays
\cite{[2001_Denniston_Yeomans]}-\cite{[2009_Parry-Jones]}.
Flexoelectric coupling in chiral and twisted nematic crystals
\cite{[1987_Patel_Meyer]}
leads to a linear rotation of the optic axis and also leads to
device applications
\cite{[2003_Blatch]}.
Flexoelectric coupling in smectic liquid crystals has been shown
to stabilize helical structures
\cite{[2001_Cepic_Zeks]}.

The purpose of this study is to calculate the values 
of the FE coefficients and other material parameters for model systems.  
The proper form of the interaction potential energy allows us 
to calculate the temperature dependence of the order parameters, 
the FE coefficients, and the elastic constants. 
The uniaxial and biaxial nematic phases are considered.

\section{Description of the system}

Let us consider a set of $N$ molecules contained in a volume $V$,
at the temperature $T$. 
The molecules are rigid blocks ($C_{2v}$ symmetry)
with three translational and three rotational degrees of freedom.
It is assumed that the molecules interact via two-body 
short-range forces
that depend on the distance between the molecules
($\vec{u} = \vec{r}_2 - \vec{r}_1 = u \vec{\Delta}$) 
and their orientations described by the three Euler angles 
$R=(\phi, \theta, \psi)$ or three orthonormal vectors
$(\vec{l},\vec{m},\vec{n})$.

The microscopic free energy $F$ for the system is given by
\begin{equation}
\label{Ftotal}
F = F_{ent} + F_{int},
\end{equation}

\begin{eqnarray}
\beta F_{ent} &=& 
\int {d\vec{r}}{dR} G(\vec{r},R)
 \{ \ln [ G(\vec{r},R) \Lambda ]-1 \},
\\
\beta F_{int} &=& 
- {\frac {1}{2}} \int {d\vec{r}_1}{dR_1}{d\vec{r}_2}{dR_2} 
G(\vec{r}_1,R_1) G(\vec{r}_2,R_2) f_{12}.
\end{eqnarray}
Here 
$f_{12}=\exp (-\beta \Phi_{12})-1 $  is the Mayer function, 
$\Phi_{12}$ the potential energy of interactions,
$dR = d\phi d\theta \sin\theta d\psi$, 
$\beta = 1/(k_{B} T)$,
and $\Lambda$ is related to the ideal gas properties.
The one-particle distribution function $G$ has the normalization
\begin{equation}
\int {d\vec{r}}{dR} G(\vec{r},R) = N.
\end{equation}
The equilibrium distribution $G$ minimizing the free energy 
(\ref{Ftotal}) satisfies the equation
\begin{equation}
\label{lnGgeneral}
\ln [G(\vec{r}_1,R_1) \Lambda]
- \int {d\vec{r}_2}{dR_2} G(\vec{r}_2,R_2) f_{12} = \mbox{const}.
\end{equation}
For the homogeneous phase $G=G_0$ does not depend on the position
and it has the form
\cite{[1997_Kapanowski]}
\begin{equation}
G_0(R) = G_0(\vec{l} \cdot \vec{L}, \vec{l} \cdot \vec{N},
\vec{n} \cdot \vec{L}, \vec{n} \cdot \vec{N}),
\end{equation}
were the orthonormal vectors $(\vec{L},\vec{M},\vec{N})$
define the biaxial nematic phase axes.
In practice we characterize the alignment not through the full 
function $G$, but by some numerical parameters - order parameters.
In the case of the biaxial nematic phase
the main order parameters are the orientational distribution averages
of the following four functions 
\cite{[1995_FKS]}:
$F_{00}^{(2)}$, $F_{02}^{(2)}$, $F_{20}^{(2)}$, and $F_{22}^{(2)}$.
We note that there are other notations
\cite{[2007_Rosso]}.
In the uniaxial nematic phase the functions
$F_{00}^{(2)}$ and $F_{02}^{(2)}$ have nonzero averages only.
The molecule alignment can be also described by order tensors $Q$ 
which are often calculated for computer simulations
\cite{[1997_Camp_Allen]}
where the molecular and laboratory axes must be distinguished.
The order tensors are defined as
\begin{eqnarray}
Q_{\alpha\beta}^{ll} & = & (3 l_{\alpha} l_{\beta} - \delta_{\alpha\beta})/2,
\\
Q_{\alpha\beta}^{mm} & = & (3 m_{\alpha} m_{\beta} - \delta_{\alpha\beta})/2,
\\
Q_{\alpha\beta}^{nn} & = & (3 n_{\alpha} n_{\beta} - \delta_{\alpha\beta})/2.
\end{eqnarray}

\section{Elastic deformations of the phase}

Orientational ordering of biaxial nematics is usually described by
the thee orthonormal vectors
\begin{equation}
\vec{L} = R_{1\alpha}\vec{e}_{\alpha},
\vec{M} = R_{2\alpha}\vec{e}_{\alpha},
\vec{N} = R_{3\alpha}\vec{e}_{\alpha}.
\end{equation}
In the homogeneous phase the vectors $(\vec{L},\vec{M},\vec{N})$
are constant in space, but in a deformed phase they depend on the
position in space.
In a continuum approach the distortion free-energy density $f_d$ 
is obtained as an expansion about an undistorted reference state
with respect to gradients of the vectors $(\vec{L},\vec{M},\vec{N})$.
The form of the $f_d$ can be derived in many alternative ways
but we use the form presented by Stallinga and Vertogen
\cite{[1994_Stallinga_Vertogen]}
(the surface terms are neglected)
\begin{eqnarray}
f_{d} & = &
{\frac{1}{2}} K_{1111} (D_{11})^{2}
+{\frac{1}{2}} K_{1212} (D_{12})^{2}
+{\frac{1}{2}} K_{1313} (D_{13})^{2} 
\nonumber \\
& & +{\frac{1}{2}} K_{2121} (D_{21})^{2}
+{\frac{1}{2}} K_{2222} (D_{22})^{2}
+{\frac{1}{2}} K_{2323} (D_{23})^{2} 
\nonumber \\
& & +{\frac{1}{2}} K_{3131} (D_{31})^{2}
+{\frac{1}{2}} K_{3232} (D_{32})^{2}
+{\frac{1}{2}} K_{3333} (D_{33})^{2} 
\nonumber \\
& & + K_{1122} D_{11} D_{22}
+ K_{1133} D_{11} D_{33}
+ K_{2233} D_{22} D_{33} 
\nonumber \\
& & + K_{1221} D_{12} D_{21}
+ K_{1331} D_{13} D_{31}
+ K_{2332} D_{23} D_{32}.
\end{eqnarray}
\begin{equation}
D_{ij} = {\frac{1}{2}} \epsilon_{jkl} R_{i\alpha} R_{k\beta}
\partial_{\alpha}R_{l\beta}.
\end{equation}
Microscopic expressions for the elastic constants $K_{ijkl}$
were derived in \cite{[1997_Kapanowski]}
and it was shown that there are 12 independent bulk constants because
\begin{equation}
K_{1221} = K_{1122},\ 
K_{1331} = K_{1133},\ 
K_{2332} = K_{2233}.
\end{equation}

\section{Flexoelectric polarization}

Liquid crystalline phases often consist of polar molecules
but in homogeneous nematic phases the average polarization is zero.
On the other hand, a phase distortion can produce a polarization
and this is called the FE effect.
In a continuum approach the FE polarization 
of the biaxial nematic phase depends on the spatial
derivatives of the vectors $(\vec{L},\vec{M},\vec{N})$
\cite{[2008_Kapanowski]}
\begin{equation}
P_{\alpha} = \sum_i (s_{ii} R_{i\alpha} \partial_{\beta} R_{i\beta}
+ b_{ii} R_{i\beta} \partial_{\beta} R_{i\alpha}).
\end{equation}
The parameters $s_{ii}$ and $b_{ii}$, $(i=1,\ 2,\ 3)$  are not unique because
if we add any constant to all of them, the polarization will not change.
The physical FE coefficients $a_i$ $(i=4,\ldots,9)$ are
\begin{eqnarray}
a_4 = s_{33}-b_{11}, & a_5 = s_{22}-b_{11}, & a_6 = s_{33}-b_{22}, 
\nonumber\\
a_7 = s_{11}-b_{22}, & a_8 = s_{22}-b_{33}, & a_9 = s_{11}-b_{33}. 
\end{eqnarray}
The coefficients satisfy the identity
\begin{equation}
a_4 - a_5 - a_6 + a_7 + a_8 - a_9 = 0.
\end{equation}
Deformations of the biaxial nematic phase connected with the
FE effect are given in Table \ref{tab1}.
In the case of the uniaxial nematic phase the FE polarization has the form
\begin{equation}
P_{\alpha} = e_1 N_{\alpha} \partial_{\beta} N_{\beta}
+ e_3 N_{\beta} \partial_{\beta} N_{\alpha}.
\end{equation}

Let us define the molecule electric dipole moment as
\begin{equation}
\mu_{\alpha} = \mu_1 l_{\alpha} + \mu_2 m_{\alpha} + \mu_3 n_{\alpha}.
\end{equation}
In the case of the molecular iteractions described below,
the FE coefficients can be expressed as follows
\begin{eqnarray}
a_4 & = & \int {d\vec{u}}{dR_1}{dR_2} f_{12} G_0(R_1)
\mu_3 n_{1z} (-u_x) (U_{2z}-W_{2x}),
\\
a_5 & = & \int {d\vec{u}}{dR_1}{dR_2} f_{12} G_0(R_1)
\mu_3 n_{1y} (-u_x)U_{2y},
\\
a_6 & = & \int {d\vec{u}}{dR_1}{dR_2} f_{12} G_0(R_1)
\mu_3 n_{1z} u_y W_{2y},
\\
a_7 & = & \int {d\vec{u}}{dR_1}{dR_2} f_{12} G_0(R_1)
\mu_3 n_{1x} u_y U_{2y},
\\
a_8 & = & \int {d\vec{u}}{dR_1}{dR_2} f_{12} G_0(R_1)
\mu_3 n_{1y} (-u_z) W_{2y},
\\
a_9 & = & \int {d\vec{u}}{dR_1}{dR_2} f_{12} G_0(R_1)
\mu_3 n_{1x} u_z (U_{2z}-W_{2x}),
\end{eqnarray}
where it is assumed that $\vec{n}$ defines the molecule
$C_{2}$ axis and
\begin{equation}
U_{\alpha} = \partial_{1} G_{0} l_{\alpha} + \partial_{3} G_{0} n_{\alpha},
\ W_{\alpha} = \partial_{2} G_{0} l_{\alpha} + \partial_{4} G_{0} n_{\alpha}.
\end{equation}

\section{Results}

We performed our calculations for the square-well potential energy
of the form
\begin{equation}
\Phi_{12}(u/ \sigma) =  
\left\{
\begin{array}{lll}
+\infty   & \mbox{for} & (u/\sigma) < 1,  \\
-\epsilon & \mbox{for} & 1 < (u/\sigma) < 2,  \\
0         & \mbox{for} & (u/\sigma) > 2,
\end{array}  
\right.
\end{equation}
were $\epsilon$ is the depth of the well and $\sigma$
depends on the molecule orientations and on the vector
$\vec{\Delta}$
\begin{eqnarray}
\label{sigma}
\sigma & = & \sigma_{0} + \sigma_{1} 
(\vec{\Delta} \cdot \vec{n}_1 - \vec{\Delta} \cdot \vec{n}_2)
+ \sigma_{2} 
\left[
(\vec{\Delta} \cdot \vec{n}_{1})^{2}+(\vec{\Delta} \cdot \vec{n}_{2})^{2}
\right] 
\nonumber\\
& & \mbox{} + \sigma_{3} 
\left[
(\vec{\Delta} \cdot \vec{l}_{1})^{2}+(\vec{\Delta} \cdot \vec{l}_{2})^{2}
\right] 
\end{eqnarray}
The parameter $\sigma_0$ defines the length scale, 
$\sigma_1$ defines the FE term,
$\sigma_2$ and $\sigma_3$ define biaxial nematic terms.
We used the density $N V_{mol}/V = 0.1$,
the molecule volume $V_{mol}$ was estimated from the mutually
excluded volume.
The FE coefficients were expressed in $\mu_i/\sigma_0^2$,
the elastic constants in $\epsilon/\sigma_0$, and
the temperature in $\epsilon/k_B$.
The parameters $\sigma_i$ are given in Table~\ref{tab2}.
The two physical systems are considered that consist of
wedge-shaped and banana-shaped molecules.

\subsection{Wedge-shaped biaxial molecules}

\begin{figure}
\begin{center}
\includegraphics{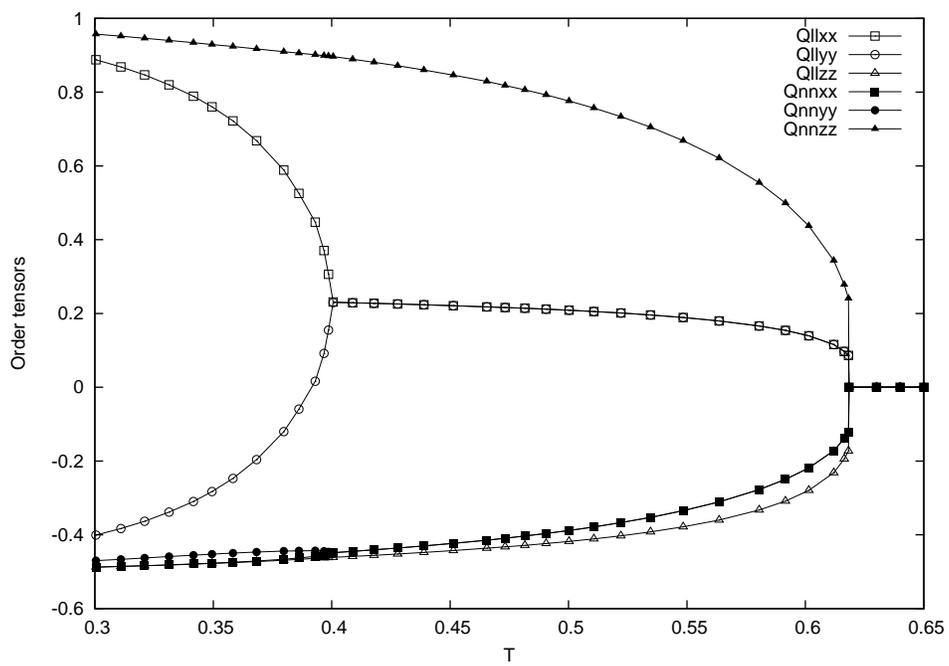}
\end{center}
\caption{
\label{fig1}
\interlinia
The temperature dependence of the order tensors
for wedge-shaped biaxial molecules.}
\end{figure}

\begin{figure}
\begin{center}
\includegraphics{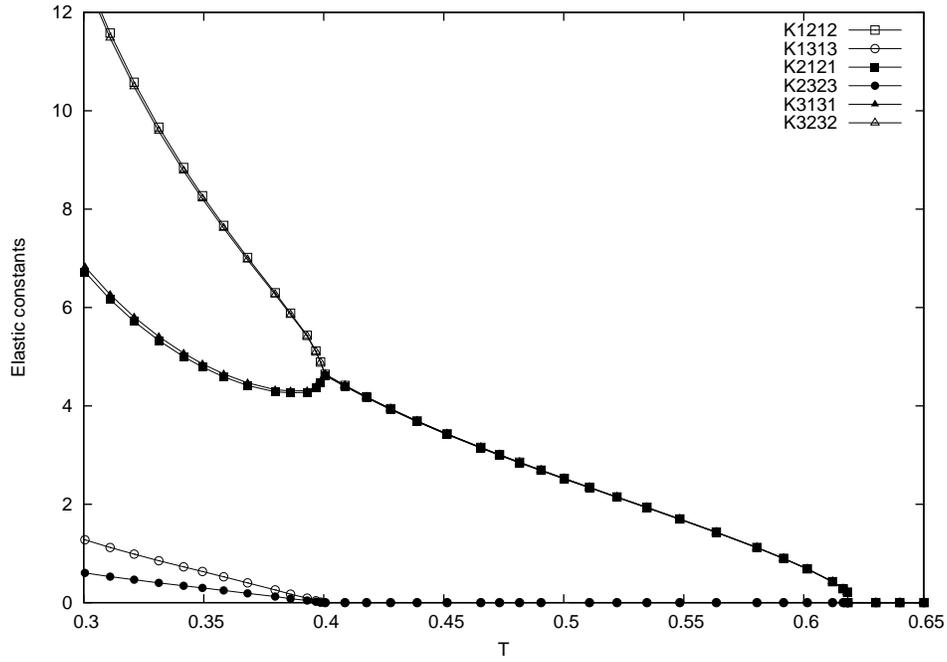}
\end{center}
\caption{
\label{fig2}
\interlinia
The temperature dependence of the elastic constants
for wedge-shaped biaxial molecules.
The squares, triangles, and circles denote deformations with
$\vec{N}$ splay, $\vec{N}$ bend, and $\vec{N}$ constant, respectively.
The empty (filled) symbols indicate the larger (smaller) parameter.}
\end{figure}

\begin{figure}
\begin{center}
\includegraphics{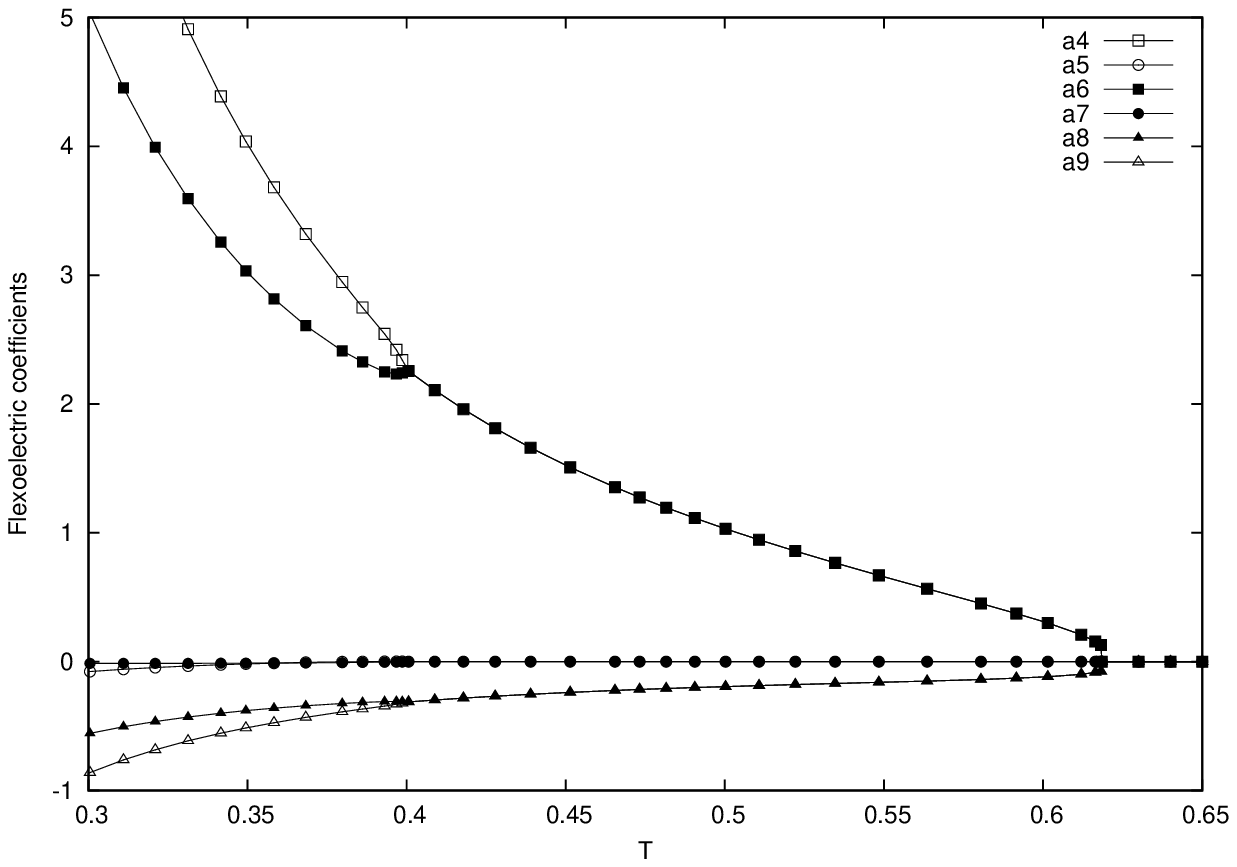}
\end{center}
\caption{
\label{fig3}
\interlinia
The temperature dependence of the flexoelectric coefficients
for wedge-shaped biaxial molecules.
Symbols have the same meaming as in Fig.~\ref{fig2}.}
\end{figure}

In the system of wedge-shaped biaxial molecules,
on decreasing the temperature we meet the first order
transition from the isotropic to the uniaxial nematic phase 
at $T_{IN}= 0.618$ and the second order transition to the biaxial
nematic phase at $T_{NB}=0.401$.
The temperature dependence of the order tensors
is presented in Fig.~\ref{fig1}.
The temperature dependence of the elastic constants and the FE
coefficients are presented in Figs.~\ref{fig2} and~\ref{fig3},
respectively.

The values of $Q_{zz}^{nn}$ show that long molecule axes
are directed along the Z axis in the whole nematic region,
whereas the values of $Q_{xx}^{ll}$ reveal the alignment
of short molecule axes along the X axis and it is
enhanced in the biaxial nematic phase.
The splay elastic constant $K_1$ splits into $K_{1212}$ and $K_{2121}$.
The bend elastic constant $K_3$ splits into $K_{3232}$ and $K_{3131}$.
Note that the equality $K_1=K_3$ is accidental and results from
neglecting order parameters $F_{\mu\nu}^{(j)}$ with $j>2$.
The splay FE coefficient $e_1$ splits into $a_4$ and $a_6$
($a_4 > a_6 > 0$).
The bend FE coefficient $-e_3$ splits into $a_8$ and $a_9$
($0 > a_8 > a_9$).
The coefficients $a_5$ and $a_7$ are small and almost always
negative.

\subsection{Banana-shaped biaxial molecules}

\begin{figure}
\begin{center}
\includegraphics{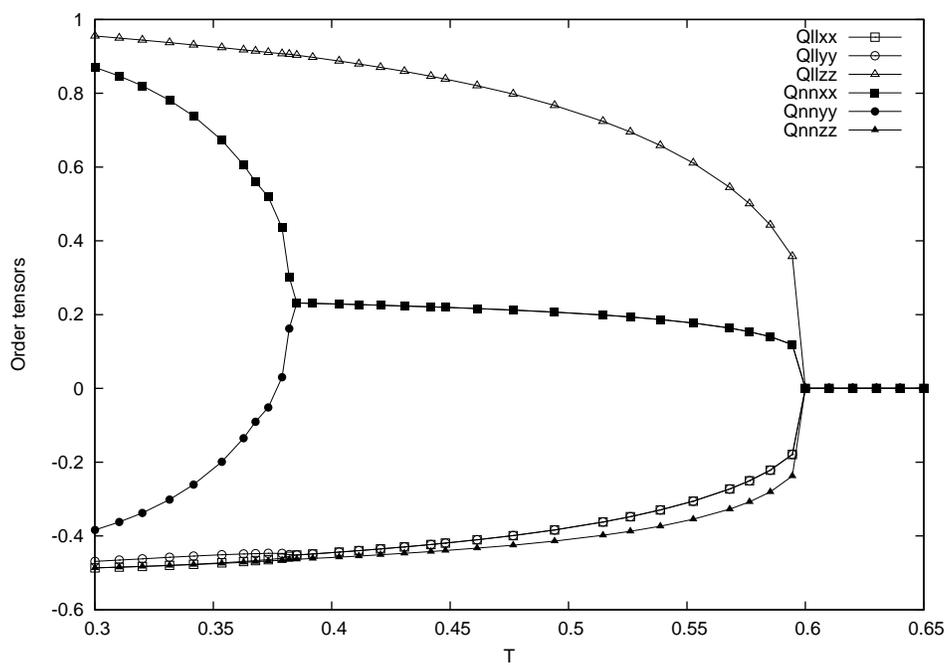}
\end{center}
\caption{
\label{fig4}
\interlinia
The temperature dependence of the order tensors
for banana-shaped biaxial molecules.}
\end{figure}

\begin{figure}
\begin{center}
\includegraphics{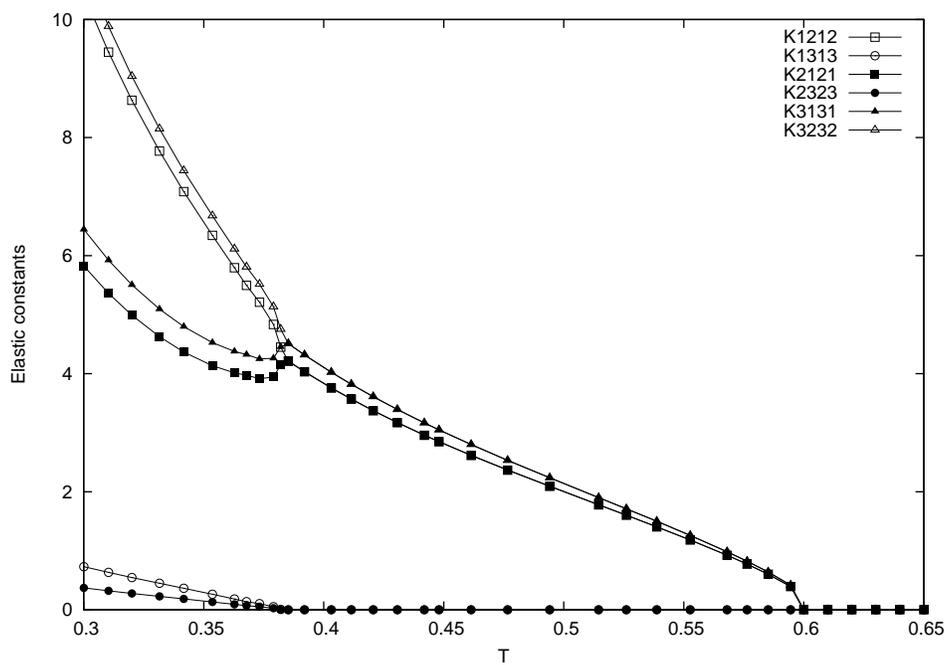}
\end{center}
\caption{
\label{fig5}
\interlinia
The temperature dependence of the elastic constants
for banana-shaped biaxial molecules.
Symbols have the same meaming as in Fig.~\ref{fig2}.}
\end{figure}

\begin{figure}
\begin{center}
\includegraphics{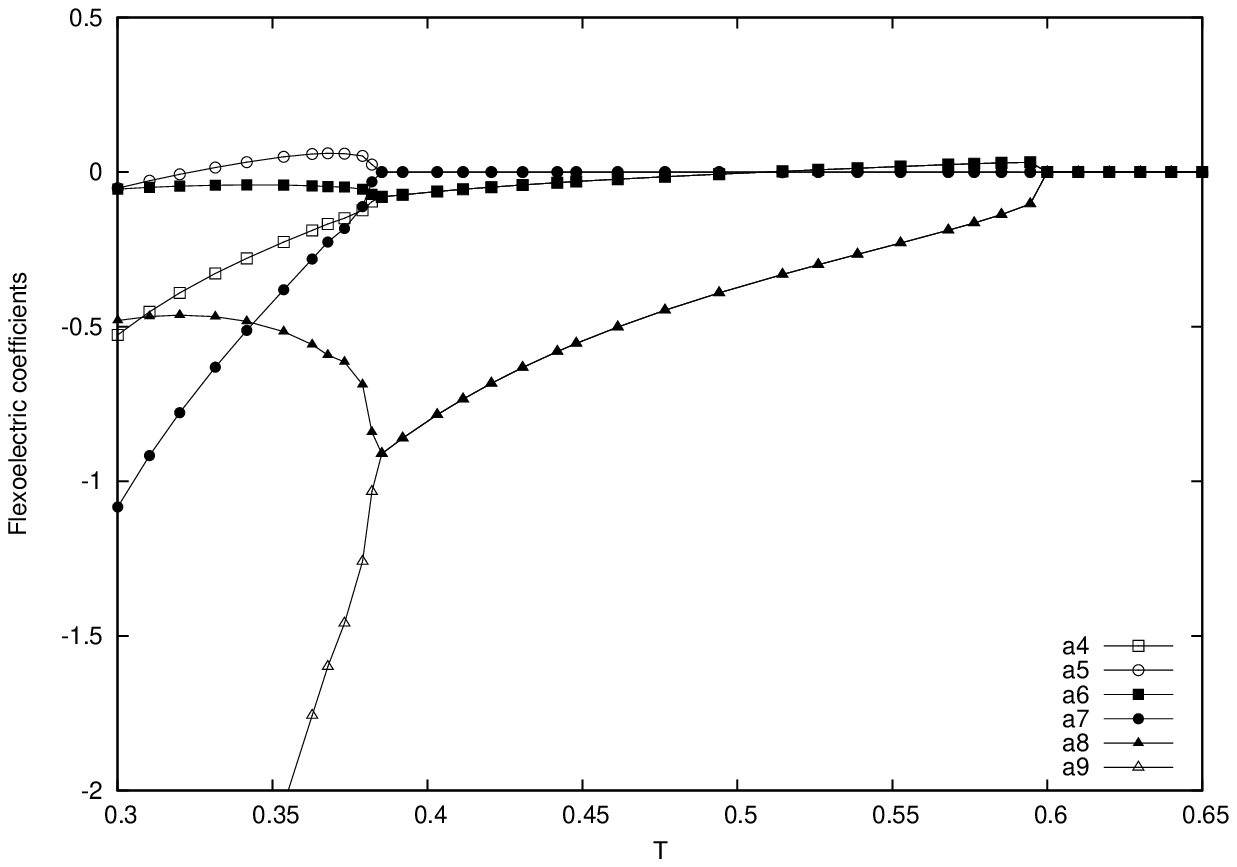}
\end{center}
\caption{
\label{fig6}
\interlinia
The temperature dependence of the flexoelectric coefficients
for banana-shaped biaxial molecules.
Symbols have the same meaming as in Fig.~\ref{fig2}.}
\end{figure}

In the system of banana-shaped biaxial molecules,
on decreasing the temperature we meet the first order
transition from the isotropic to the uniaxial nematic phase 
at $T_{IN}= 0.595$ and the second order transition to the biaxial
nematic phase at $T_{NB}=0.382$.
The temperature dependence of the order tensors
is presented in Fig.~\ref{fig4}.
The temperature dependence of the elastic constants and the FE
coefficients are presented in Figs.~\ref{fig5} and~\ref{fig6},
respectively.

According to the values of $Q_{zz}^{ll}$ long molecule axes
are directed along the Z axis in the whole nematic region,
whereas the values of $Q_{xx}^{nn}$ show the alignment
of short molecule axes along the X axis and it is
enhanced in the biaxial nematic phase.
The behaviour of the elastic constants is similar to the case
of the wedge-shaped molecules because the FE term is small
in both cases.
The bend FE coefficient $-e_3$ splits into $a_8$ and $a_9$
($0 > a_8 > a_9$).
The splay FE coefficient $e_1$ is smaller then $e_3$ and
it splits into $a_4$ and $a_6$.
The coefficients $a_5$ and $a_7$ are again small but comparable
with $a_4$ and $a_6$. The sign of some coefficient can change
on changing the temperature.

\section{Conclusions}

In this paper, the statistical theory was used to study
the temperature dependence of the order parameters, elastic
constants, and FE coefficients of biaxial nematic liquid crystals.
In order to calculate these macroscopic parameters one needs
the one-particle distribution function and the potential energy
of molecular interactions. 
The two physical systems were considered.
The splittings of the FE coefficients and the elastic constants
were obtained at the uniaxial-biaxial nematic transition.
New small FE coefficients appeared at the transition.
The ordering of the splited FE coefficients
corresponds to the ordering of the splited elastic constants.

The FE coefficients were proportional to the dipole moment component
parallel to the molecule $C_{2v}$ symmetry axis.
This was the result of the interactions potential symmetry.
The beaviour of the main FE coefficients, $e_1$ for the wegde-shaped
molecules and $e_3$ for the banana-shaped molecules, is clear
and it is in the agreement with previous studies 
\cite{[2008_Kapanowski_OER]}.
On the other hand, it seems that other FE coefficients should
be interpreted with caution. It is possible that higher order 
parameters can have a significant contribution.

At present stage, the direct comparison between the theory 
and the experiment in not possible because to our knowledge
the FE coefficients have not been measured for the biaxial
nematic phase. What is more, even for the uniaxial nematic phase
the experimental data are still scarce and sometimes
contradictory~\cite{[2001_Petrov]}.
However, when biaxial nematic phases become more widespread,
the presented theory will be helpful in practical applications.

\section*{Acknowledgements}

The author is grateful to J. Spa{\l}ek for his support
and discussions.

\begin{table}
\caption{
\label{tab1}
Deformations of the biaxial nematic phase connected with the
FE effect. The corresponding elastic constants and the FE coefficients
are given, the values for the uniaxial nematic phase
are in parentheses.}
\begin{center}
\begin{tabular}{ccc}
\hline\hline
Deformation & Elastic constant & FE coefficient
\\ \hline
$\vec{N}$ splay, $\vec{L}$ bend & $K_{1212}$ ($K_1$) & $a_4$ ($e_1$) \\
$\vec{M}$ splay, $\vec{L}$ bend & $K_{1313}$ (0)     & $a_5$ (0) \\
$\vec{N}$ splay, $\vec{M}$ bend & $K_{2121}$ ($K_1$) & $a_6$ ($e_1$) \\
$\vec{L}$ splay, $\vec{M}$ bend & $K_{2323}$ (0)     & $a_7$ (0) \\
$\vec{M}$ splay, $\vec{N}$ bend & $K_{3131}$ ($K_3$) & $a_8$ ($-e_3$) \\
$\vec{L}$ splay, $\vec{N}$ bend & $K_{3232}$ ($K_3$) & $a_9$ ($-e_3$) \\
\hline\hline
\end{tabular}
\end{center}
\end{table}

\begin{table}
\caption{
\label{tab2}
Parameters $\sigma_i$ used in calculations.}
\begin{center}
\begin{tabular}{ccccccc}
\hline\hline
Molecules & $\sigma_1/\sigma_0$ & $\sigma_2/\sigma_0$ & $\sigma_3/\sigma_0$ 
& Long axis & Short axis & $C_{2}$ axis\\
\\ \hline
wedge-like  & $0.2$  & $0.5$  & $-0.4$ & $\vec{n}$ & $\vec{l}$ & $\vec{n}$ \\
banana-like & $0.2$  & $-0.4$  & $0.5$ &  $\vec{l}$ & $\vec{n}$ & $\vec{n}$ \\
\hline\hline
\end{tabular}
\end{center}
\end{table}

\begin{thebibliography}{99}

\bibitem{[1970_Freiser]}
M. J. Freiser, "Ordered states of a nematic liquid",
{\em Phys. Rev. Lett.} {\bf 24}, 1041-1043 (1970).

\bibitem{[1980_Yu_Saupe]}
L. J. Yu and A. Saupe, "Observation of a biaxial nematic phase
in potassium-1-decanol-water mixtures",
{\em Phys. Rev. Lett.} {\bf 45}, 1000-1003 (1980).

\bibitem{[2004_Severing_Saalwachter]}
K. Severing and K. Saalwachter, "Biaxial nematic phase in 
a thermotropic liquid crystalline side-chain polymer",
{\em Phys. Rev. Lett.} {\bf 92}, 125501 (2004).

\bibitem{[2004_Madsen]}
L. A. Madsen, T. J. Dingemans, M. Nakata, and E. T. Samulski,
"Thermotropic biaxial nematic liquid crystals",
{\em Phys. Rev. Lett.} {\bf 92}, 145505 (2004).

\bibitem{[2004_Acharya]}
B. R. Acharya, A. Primak, and S. Kumar, 
"Biaxial nematic phase in bent-core thermotropic mesogens",
{\em Phys. Rev. Lett.} {\bf 92}, 145506 (2004).

\bibitem{[2004_Merkel]}
K. Merkel, A. Kocot, J. K. Vij, R. Korlacki, G. H. Mehl,
and T. Meyer, 
"Thermotropic biaxial nematic phase in liquid crystalline
organo-siloxane tetrapodes",
{\em Phys. Rev. Lett.} {\bf 93}, 237801 (2004).

\bibitem{[2005_Longa_Grzybowski]}
L. Longa, P. Grzybowski, S. Romano, and E. Virga,
"Minimal coupling model of the biaxial nematic phase",
{\em Phys. Rev.} {\bf E 71}, 051714 (2005).

\bibitem{[2008_Berardi]}
R. Berardi, L. Muccioli, S. Orlandi, M. Ricci, and C. Zannoni,
"Computer simulations of biaxial nematics",
{\em J. Phys.: Condens. Matter} {\bf 20}, 463101 (2008).

\bibitem{[2007_Longa]}
L. Longa, G. Paj±k, and T. Wydro, 
"Stability of biaxial nematic phase for systems with
variable molecular shape anisotropy",
{\em Phys. Rev.} {\bf E 76}, 011703 (2007).

\bibitem{[1969_Meyer]}
R. B. Meyer, "Piezoelectric effects in liquid crystals", 
{\em Phys. Rev. Lett.} {\bf 22}, 918-921 (1969).

\bibitem{[1977_Prost_Marcerou]}
J. Prost and J. P. Marcerou, 
"On the microscopic interpretation of flexoelectricity",
{\em J. Phys. (Paris)} {\bf 38}, 315-324 (1977).

\bibitem{[2008_Kapanowski]}
A. Kapanowski, "Flexoelectric polarization in the biaxial nematic phase", 
{\em Phys. Rev.} {\bf E 77}, 052702 (2008).

\bibitem{[2001_Denniston_Yeomans]}
C. Denniston and J. M. Yeomans, 
"Flexoelectric surface switching of bistable nematic devices",
{\em Phys. Rev. Lett.} {\bf 87}, 275505 (2001).

\bibitem{[2002_Davidson_Mottram]}
A. J. Davidson and N. J. Mottram, 
"Flexoelectric switching in a bistable nematic device",
{\em Phys. Rev.} {\bf E 65}, 051710 (2002).

\bibitem{[2009_Parry-Jones]}
L. A. Parry-Jones, R. B. Meyer, and S. J. Elston,
"Mechanisms of flexoelectric switching in a zenithally
bistable nematic device",
{\em J. Appl. Phys.} {\bf 106}, 014510 (2009).

\bibitem{[1987_Patel_Meyer]}
J. S. Patel and R. B. Meyer, 
"Flexoelectric electro-optics of a cholesteric liquid crystal",
{\em Phys. Rev. Lett.} {\bf 58}, 1538-1540 (1987).

\bibitem{[2003_Blatch]}
A. E. Blatch, M. J. Coles, B. Musgrave, and H. J. Coles,
"Flexoelectric liquid crystal bimesogens",
{\em Mol. Cryst. Liq. Cryst.} {\bf 401}, 161-169 (2003).

\bibitem{[2001_Cepic_Zeks]}
M. Cepic and B. Zeks, 
"Flexoelectricity and Piezoelectricity: The Reason for the Rich 
Variety of Phases in Antiferroelectric Smectic Liquid Crystals",
{\em Phys. Rev. Lett.} {\bf 87}, 085501 (2001).


\bibitem{[1997_Kapanowski]}
A. Kapanowski, "Statistical theory of elastic constants of biaxial
nematic liquid crystals",
{\em Phys. Rev.} {\bf E 55}, 7090-7104 (1997).

\bibitem{[1995_FKS]}
M. Fialkowski, A. Kapanowski, K. Sokalski, 
"Microscopic approach to theory of biaxial nematic liquid crystals",
{\em Mol. Cryst. Liq. Cryst.} {\bf 265}, 371-385 (1995).

\bibitem{[1994_Stallinga_Vertogen]}
S. Stallinga, G. Vertogen, "Theory of orientational elasticity",
{\em Phys. Rev.} {\bf E 49}, 1483-1495 (1994).

\bibitem{[2007_Rosso]}
R. Rosso, "Orientational order parameters in biaxial nematics:
Polymorfic notation",
{\em Liq. Cryst.} {\bf 34}, 737-748 (2007).

\bibitem{[1997_Camp_Allen]}
P. J. Camp and M. P. Allen, 
"Phase diagram of the hard biaxial ellipsoid fluid",
{\em J. Chem. Phys.} {\bf 106}, 6681 (1997).

\bibitem{[2008_Kapanowski_OER]}
A. Kapanowski, "Flexoelectric effect modelling",
{\em Opto-Electron. Rev.} {\bf 16}, 9-12 (2008).

\bibitem{[2001_Petrov]}
A. G. Petrov, Measurements and interpretation of flexoelectricity",
in {\em Physical properties of liquid crystals: nematics}, 
pp. 251-264, edited by D. Dunmur, A. Fukuda, and G. L. Luckhurst,
INSPEC, The Institution of Electrical Engineers, London, 2001.

\end{thebibliography}
\end{document}